\documentclass[a4paper,11pt]{article}
\pdfoutput=1

\usepackage{jheppub} 
\usepackage{graphicx}
\usepackage{hyperref}
\usepackage[utf8]{inputenc} 
\usepackage{amsmath}
\usepackage{amssymb}
\usepackage{float}
\usepackage{comment}
\usepackage{slashed}
\usepackage[normalem]{ulem}
\usepackage{breqn}
\usepackage{caption}
\usepackage{soul}
\usepackage{cancel}
\usepackage{subcaption}
\usepackage{booktabs}
\usepackage{mathtools,braket,blkarray}
\usepackage[usenames,dvipsnames]{color}
\usepackage[colorinlistoftodos]{todonotes}
\usepackage[normalem]{ulem}
\usepackage{multirow}
\usepackage[colorinlistoftodos]{todonotes}
\usepackage{orcidlink}

 \newcommand {\ignore}[1]{}

\usepackage{ulem}

\def\21{$\mathrm{SU(2)_L \otimes U(1)_Y}$}

\def\bea{\begin{eqnarray}}
\def\eea{\end{eqnarray}}
\def\beq{\begin{equation}}
\def\eeq{\end{equation}}

\def\nn{\nonumber}

\newcommand{\lsim}{
\mathrel{\hbox{\rlap{\hbox{\lower4pt\hbox{$\sim$}}}\hbox{$<$}}}}
\newcommand{\gsim}{
\mathrel{\hbox{\rlap{\hbox{\lower4pt\hbox{$\sim$}}}\hbox{$>$}}}}

\title{\boldmath Self-Interacting Sterile Neutrino Cold Dark Matter: Resonant Production Mechanism in the Early Universe}

\author[a]{{Eung Jin Chun,}\orcidlink{0000-0003-4786-313X}}
\affiliation[a]{Korea Institute for Advanced Study, Seoul, 02455, Korea}
\emailAdd{ejchun@kias.re.kr}

\author[b]{Kenji Kadota,}
\affiliation[b]{School of Fundamental Physics and Mathematical Sciences, Hangzhou Institute for Advanced Study, University of Chinese Academy of Sciences (HIAS-UCAS), Hangzhou, 310 024, China}

\author[c,d]{{Seokhoon Yun}\orcidlink{0000-0002-7960-3933}}
\emailAdd{seokhoon.yun@knu.ac.kr}
\affiliation[c]{Department of Physics, Kyungpook National University, Daegu 41566, Korea}
\affiliation[d]{Center for Theoretical Physics of the Universe, Institute for Basic Science (IBS), Daejeon, 34126, Korea}

 \abstract{
Sterile neutrinos are well-motivated dark matter candidates, but their conventional production through active--sterile mixing is tightly constrained by X-ray searches and structure-formation observations. 
We propose a distinct production mechanism operating entirely within a sterile sector: two sterile neutrinos, $N_1$ and $N_2$, coupled to a singlet scalar $\phi$, with $N_1$ the dark matter candidate and $N_2$ held in equilibrium through frequent scattering induced by its scalar interaction. Thermal self-energies induced by the $N_2$ and $\phi$ backgrounds generate both a temperature-dependent mass splitting and an off-diagonal mixing between  $N_1$ and $N_2$. As the Universe cools, the in-medium levels can undergo a level crossing, leading to resonantly enhanced conversion of the thermal $N_2$ population into  $N_1$.
We formulate the conversion using a density-matrix kinetic equation that consistently incorporates coherent $N_1$--$N_2$ conversion, collisional decoherence, and thermal repopulation of $N_2$.
For a narrow resonance, the integrated conversion probability admits a simple analytic form that coincides with the Landau--Zener result, despite the underlying collisionally damped dynamics.
In the weak-conversion regime relevant for freeze-in, this correspondence provides a robust analytic description of the resonant production.
We derive the resulting dark matter abundance and identify the conditions for cosmological stability of  $N_1$ and for resonant conversion to dominate over direct scattering and decay production. The resulting relic abundance scales as $Y_1\propto g_{12}^2 g_{22}M_{\rm Pl}/m_1$, making the dark matter energy density approximately independent of $m_1$. This mechanism provides a new route to sterile-neutrino dark matter that does not require appreciable active--sterile mixing.
}

\begin{document} 
\preprint{CTPU-PTC-26-23}
\maketitle
\flushbottom

\section{Introduction}
\label{sec:intro}


The observation of neutrino oscillations provides direct evidence for physics beyond the Standard Model (SM), requiring nonzero neutrino masses and mixing.
One of the simplest extensions capable of accommodating neutrino masses is the introduction of gauge-singlet fermions, conventionally referred to as sterile or right-handed neutrinos.
Depending on their masses and interactions, sterile neutrinos can play a variety of roles in particle physics and cosmology, including the generation of active-neutrino masses through the seesaw mechanism, the production of the baryon asymmetry, and the origin of dark matter.
In particular, a sufficiently long-lived sterile neutrino with sufficiently weak interactions provides a minimal dark matter candidate whose phenomenology connects particle physics, astrophysics, and cosmology~\cite{Drewes:2016upu}. 

A particularly attractive feature of sterile-neutrino dark matter is that its abundance can, in principle, be generated solely through mixing with active neutrinos.
In the original Dodelson--Widrow mechanism~\cite{Dodelson:1993je}, active neutrinos in the primordial plasma undergo incoherent conversion into sterile states through their vacuum mixing.
The same mixing responsible for production also induces the radiative decay of the sterile neutrino, leading to a monochromatic photon signal that can be searched for in X-ray observations.
At the same time, the non-thermal momentum distribution of the produced sterile neutrinos affects structure formation on small scales.
The combination of X-ray searches, phase-space considerations, and structure-formation constraints has therefore provided powerful tests of this minimal scenario.
In particular, the simplest Dodelson--Widrow realization in which sterile neutrinos constitute the entire dark matter abundance cannot simultaneously satisfy the relic-density requirement and the existing X-ray and structure-formation limits~\cite{Abazajian:2017tcc,TerolCalvo:2026nlr}.

This tension has motivated a broad range of alternative production mechanisms.
A well-known possibility is resonant active--sterile conversion in the presence of a primordial lepton asymmetry, as in the Shi--Fuller mechanism~\cite{Shi:1998km}.
The modified matter potential can strongly enhance active--sterile mixing over a limited range of temperatures and momenta, resulting in a colder dark matter spectrum than in non-resonant production.
Other possibilities include production from the decays or annihilations of additional particles, freeze-in through feeble interactions, and scenarios involving new interactions within the active or sterile sector, and so on \cite{Dasgupta:2013zpn,Kadota:2017hzy,Jeong:2018yts,DeGouvea:2019wpf,Johns:2019cwc,Bringmann:2021tjr,Chen:2022kal,Balantekin:2023jlg,Astros:2023xhe,Fuller:2024noz,Dev:2025sah}.
Such constructions illustrate that the cosmological abundance of sterile-neutrino dark matter need not be directly tied to its vacuum mixing with active neutrinos. 

In this work, we explore another possibility based on the modification of particle propagation in a thermal medium.
Interactions with a thermal background change the dispersion relations of propagating states and, in a multi-state system, can also induce off-diagonal contributions to the effective Hamiltonian.
The resulting in-medium mixing can differ substantially from vacuum mixing and may become resonantly enhanced when the dispersion relations of two states approach each other.
This motivates us to investigate whether thermal effects entirely within a sterile sector can provide an efficient production mechanism for sterile-neutrino dark matter, whose mass may significantly exceed the conventional keV range.

We consider two sterile neutrinos, $N_1$ and $N_2$, coupled to a singlet scalar $\phi$, where $N_1$ serves as the dark matter candidate while $N_2$ remains in thermal equilibrium.
The thermal populations of $N_2$ and $\phi$ induce self-energy corrections that modify the sterile-neutrino dispersion relations and generate an effective $N_1$--$N_2$ mixing.
As the Universe cools, the corresponding thermal corrections evolve, and for $m_1>m_2$ the in-medium dispersion relations can cross at a finite temperature.
Around this level crossing, the effective mixing is resonantly enhanced, allowing a fraction of the thermal $N_2$ population to be converted into the otherwise inert $N_1$ state.
Unlike conventional sterile-neutrino dark matter production, both the mixing and the resonance in our scenario originate from interactions within the sterile sector rather than from active--sterile vacuum mixing and the SM matter potential.

The frequent collisions that maintain $N_2$ in thermal equilibrium also induce collisional decoherence during the resonant conversion.
Consequently, the conversion cannot, in general, be described as an isolated coherent two-state oscillation.
This should be contrasted with scenarios based on coherent Landau--Zener evolution, such as Ref.~\cite{Anisimov:2008gg}.
We therefore formulate the production process using a density-matrix kinetic equation that consistently incorporates coherent $N_1$--$N_2$ conversion together with production, absorption, and collisional decoherence of $N_2$.
In the parameter region of interest, the resonance is narrow compared with the Hubble timescale, allowing the microscopic conversion dynamics to be treated locally while cosmological expansion controls the slow evolution of the thermal background.
Remarkably, despite the collisionally damped microscopic dynamics, the conversion probability integrated over a narrow resonance takes the same functional form as the Landau--Zener result within the quasi-stationary treatment.
Moreover, in the perturbative weak-conversion regime relevant for freeze-in, the leading integrated result remains valid even when the local quasi-stationary approximation is relaxed, provided that the level crossing is approximately linear and the relevant background quantities vary slowly across the transition interval.


Using this framework, we derive an analytic expression for the $N_1$ abundance generated by resonant thermal mixing.
A notable feature of the resulting comoving $N_1$ abundance, $Y_1\equiv n_1/s$, is [see Eq.~(\ref{eq:Coupling}) for notations]
\bea
Y_1^{\rm mix}
\propto
g_{12}^2g_{22}\frac{M_{\rm P}}{m_1}\,,
\eea
so that the corresponding comoving dark matter energy density $m_1Y_1^{\rm mix}$ becomes approximately independent of $m_1$, apart from corrections associated with the sterile-sector mass hierarchy and the relativistic degrees of freedom.
We further determine the conditions under which the produced $N_1$ remains cosmologically stable and resonant conversion dominates over competing direct production processes, including scalar decays and $2\rightarrow2$ scatterings.
These results identify a viable parameter region in which the observed dark matter abundance is predominantly generated through resonant thermal $N_1$--$N_2$ conversion.


The remainder of this paper is organized as follows.
In Sec.~\ref{sec:setup}, we introduce the sterile-sector model and discuss the thermalization of $N_2$ and $\phi$.
Then, we derive the finite-temperature corrections to the sterile-neutrino dispersion relations and the induced $N_1$--$N_2$ mixing.
In Sec.~\ref{sec:darkmatter}, we formulate the density-matrix evolution and calculate the production rate of $N_1$ through resonant thermal mixing.
We then derive the resulting dark matter abundance and discuss the conditions for stability and the dominance of resonant production.
Finally, we summarize our results and discuss their implications in Sec.~\ref{sec:conc}.


\section{Thermal Masses and Mixing of Sterile Neutrinos}
\label{sec:setup}

In our setup, the sterile neutrino sector consists of two mass-eigenstate fields, $N_1$ and $N_2$, with vacuum masses $m_{1}$ and $m_{2}$, respectively.
The $N_1$ state is assumed to be inert with respect to the SM sector and serves as a dark matter (DM) candidate, while $N_2$ interacts efficiently with SM particles, which may play a role in generating their masses.

To incorporate self-interactions within the sterile neutrino sector, we consider a singlet scalar field $\phi$ that couples to the sterile Majorana neutrinos ($N^c_i =N_i$) through Yukawa interactions of the form
\bea
-\mathcal{L}_{\rm RHN} = \sum_{ij}{1\over2} g_{ij} \phi \bar{N}_i N_j + \sum_i {1\over2} m_i \bar N_i N_i, 
\label{eq:Coupling}
\eea
where $g_{ij}$ denote the symmetric Yukawa couplings which are assumed to be real for simplicity. 

We are interested in the situation that the production of $N_1$ DM proceeds through its effective mixing with $N_2$, which is mediated by coherent forward scatterings in the thermal plasma.
This thermal mixing enables a partial conversion of the thermal abundance of $N_2$ into $N_1$ in a non-thermal (freeze-in) way. 

The mass of active neutrinos $m_\nu$ arises from the Yukawa interaction between $N_2$ and the SM Higgs field, $y_\nu \, H\bar{N}_2 \nu$, with the coupling estimated as $y_\nu \sim \sqrt{m_\nu m_{2}}/v_{\rm EW}$, where $v_{\rm EW}$ is the electroweak vacuum expectation value.
For phenomenologically relevant neutrino masses $m_\nu \sim 0.1$ eV, this interaction alone is too weak to generate sizable thermal mass correction while maintaining thermalization of $N_2$ at temperatures well above its mass.

To ensure that $N_2$ is thermalized during the relevant cosmological epoch, we assume that the scalar $\phi$ itself is in thermal equilibrium with the SM plasma, for instance, through its interaction with the Higgs field: $\phi+\phi \leftrightarrow H +\bar H$.
Once thermalized, $\phi$ efficiently brings $N_2$ into equilibrium through the process $\phi + \phi \leftrightarrow N_2 + \bar{N}_2$, at temperatures $T \lesssim g_{22}^4 m_{\rm pl}$ requiring the production rate larger than the Hubble rate: $\Gamma_2 >H$ with $\Gamma_2 \sim g_{22}^4 T/4\pi$ and $H=1.66 \sqrt{g_*} T^2/m_{\rm pl}$.
We will take $g_{22}$ to be sufficiently large so that this condition is easily satisfied, ensuring a thermal population of $N_2$ during the epoch relevant for $N_1$ DM production.



\begin{figure}[t!]
\centering
 \includegraphics[width=.8\linewidth]{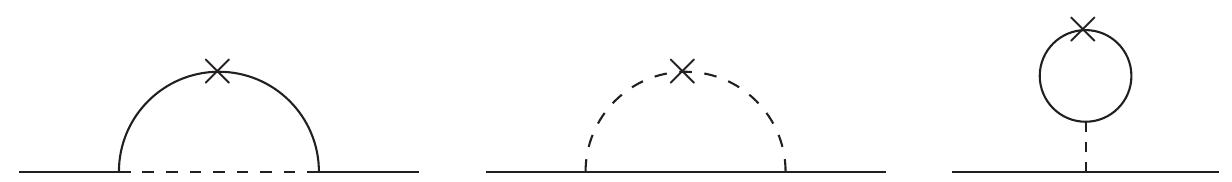}
 \caption{Thermal loop diagrams contributing to the dispersion relation of sterile neutrinos. Solid lines represent sterile neutrinos $N_i$, dashed lines denote the scalar $\phi$, and crosses indicate $N_2$ or $\phi$ particles in the thermal bath.} 
 \label{fig:thermalloop}
\end{figure}

For particles propagating in the thermal background of $\phi$ and $N_2$, their coherent forward scattering, mediated by the self-interactions in Eq.~\eqref{eq:Coupling}, induces thermal corrections to the dispersion relations \cite{Weldon:1982bn,Nieves:1989ez,Quimbay:1995jn}.
The leading order diagrams contributing to these corrections are depicted in Fig.~\ref{fig:thermalloop}, where solid and dashed lines represent sterile neutrinos $N_{i}$ and the scalar $\phi$, respectively.
Crosses on the lines indicate $N_2$ (solid) or $\phi$ (dashed) particles in the thermal bath, acting as scattering targets for coherent forward scattering.

For the sake of simplicity, we assume a symmetric thermal background in which particle and antiparticle occupation numbers are identical.
This assumption is typically relevant in the high-temperature and low chemical potential regime of the early universe.
We therefore focus on the relativistic limit (i.e., $T\gg m_{i}\,, m_\phi$) and consider external momenta around the temperature scale as the typical value.

The first diagram in Fig.~\ref{fig:thermalloop} depicts the external $N_i$ states propagating in the thermal background of $N_2$ via $\phi$ exchange.
Its contribution leads to the effective Lagrangian
\bea
\mathcal{L}_{N_2{\rm -loop}} = 2 \int \frac{d^3 \vec{k}}{\left(2\pi\right)^3 2 E_k}f_{2}\,\left(-\frac{g_{i2}g_{2j}}{2}\right)\bar{N}_i \left[\frac{\cancel{k}-m_{2}}{\left(p + k\right)^2-m_\phi^2} - \frac{\cancel{k}+m_{2}}{\left(p - k\right)^2-m_\phi^2}\right]N_j \, ,
\eea
where $p$ and $k$ denote the four-momenta of the external states and thermal $N_2$, respectively, and $f_{2} = (e^{E_k/T}+1)^{-1}$ is the Fermi-Dirac distribution.
The factor of $1/2$ accounts for the spin sum average of the thermal $N_2$ states (and the prefactor $2$ counts on the total degree of freedom).
In the relativistic regime, where the product $(p\cdot k)$ dominates the scalar propagator, the expression simplifies to
\bea
\left. \mathcal{L}_{N_2{\rm -loop}} \right|_{\rm rel} \approx \left(-g_{i2}g_{2j}\right) \,  \bar{N}_i \left[ \frac{1}{2 E_p} \gamma^0 V_{2} - \frac{1}{2 |\vec{p}|} \vec{\gamma}\cdot\hat{p} \left(V_{2} - \frac{T^2}{24}\right)\right]N_j \, ,
\label{eq:N2Loop}
\eea
where $E_p$ and $\hat{p}=\vec{p}/|\vec{p}|$ are the energy and the momentum direction of the external $N_i$ state, respectively.
The potential factor $V_{2}$ is given by~\cite{Quimbay:1995jn}
\bea
V_{2} = \frac{1}{4\pi^2}  \int \frac{|\vec{k}|^2}{E_k} d|\vec{k}| f_{2} \left(\frac{E_k}{|\vec{k}|} \log\frac{E_k + |\vec{k}|}{E_k - |\vec{k}|}\right) \approx
\frac{T^2}{48}\log\frac{T^2}{m_{2}^2}  \,. 
\eea
As we will see explicitly below, $V_2$ enters the temporal and spatial components of the equation of motion in the same way and therefore cancels from the effective mass-squared dispersion relation in the relativistic limit.

The second diagram in Fig.~\ref{fig:thermalloop} involves coherent scatterings with thermal $\phi$ particles.
The corresponding contribution to the effective Lagrangian reads 
\bea
\left. \mathcal{L}_{\phi{\rm -loop}} \right|_{\rm rel} & = &\int \frac{d^3 \vec{l}}{\left(2\pi\right)^3 2 E_l} f_{\phi}\,\left(-\sum_k g_{ik}g_{kj}\right)\bar{N}_i \left[\frac{\cancel{p}+\cancel{l}-m_k}{\left(p + l\right)^2-m_k^2} + \frac{\cancel{p}-\cancel{l}+m_{k}}{\left(p - l\right)^2-m_k^2}\right]N_j \, , \nonumber\\
& \approx & \left(-2\sum_k g_{ik}g_{kj}\right) \, \bar{N}_i \left[ \frac{1}{2 E_p} \gamma^0 V_{\phi} - \frac{1}{2 |\vec{p}|} \vec{\gamma}\cdot\hat{p} \left(V_{\phi} - \frac{T^2}{24}\right)\right]N_j \, ,
\label{eq:phiLoop}
\eea
where $E_l$ and $\vec{l}$ denote the energy and momentum of the external $\phi$, respectively, and $f_\phi =  (e^{E_l/T}-1)^{-1}$ is the Bose-Einstein distribution.
The potential $V_\phi$ reads
\bea
V_{\phi} = \frac{1}{4\pi^2}  \int \frac{|\vec{l}|^2}{E_l} d|\vec{l}| f_{2} \left(\frac{E_l}{|\vec{l}|} \log\frac{E_l + |\vec{l}|}{E_l - |\vec{l}|}\right) \approx \frac{T^2}{ 48}\log\frac{T^2}{m_{\phi}^2}  \,. 
\eea
Analogously to $V_2$, the potential $V_\phi$ appears symmetrically in the temporal and spatial components and does not contribute to the effective mass-squared dispersion relation at leading order in the relativistic limit.

The third diagram in Fig.~\ref{fig:thermalloop} is a tadpole involving the thermal $N_2$ loop and contributes as a mass term
\bea
\left. \mathcal{L}_{{\rm tad}} \right|_{\rm rel} & = & \int \frac{d^3 \vec{k}}{\left(2\pi\right)^3 2 E_k}f_{N_2}\left(-\frac{g_{ij}g_{22}}{2}\right)\,\bar{N}_iN_j \left(\frac{1}{m_\phi^2}\right) \left({\rm tr}\left[\cancel{k}+m_{2}\right]-{\rm tr}\left[\cancel{k}-m_{2}\right]\right) \, \nn\\
& \simeq & - 4 m_{2} \left(\frac{g_{ij}g_{22}}{m_\phi^2}\right)  \frac{T^2}{24}\,\bar{N}_iN_j  \, .
\label{eq:N2tadpole}
\eea
Unlike the loop corrections in Eqs.~\eqref{eq:N2Loop} and \eqref{eq:phiLoop}, the tadpole contribution therefore acts as a direct correction to the scalar mass term.

The thermal bath also modifies the propagation of $\phi$ itself.
In particular, the thermal sterile-neutrino loop generates an effective (static) potential for $\phi$ of the form
\bea
\mathcal{L}_{N{\rm -loop}}\approx -\left(\sum_i g_{i2} g_{2i}\right){T^2 \over 6}\,\phi^2 . 
\eea
This effectively screens the scalar and renders the tadpole contribution temperature-independent in the high-temperature regime, where $m_\phi$ is sub-dominant in the $\phi$ dispersion (i.e., $m_\phi^2 \ll \left(\sum_i g_{i2}^* g_{2i}\right)T^2/3$).

Combining the contributions above, the equation of motion for the sterile-neutrino system can be written as
\bea
 \left[ \left(E \delta_{ij} - \frac{V_{ij}}{2E}\right) \gamma^0 - \left(|\vec{p}|\delta_{ij} - \frac{V_{ij} - \Delta m_{ij}^2}{2|\vec{p}|}\right)\vec \gamma \cdot \hat p -  \left(m_{i} \delta_{ij}+\delta m_{ij}\right) \right] N_j =  0 \,,
\label{eq:DispersionPotential}
\eea
where
\bea
 V_{ij} &=& g_{i2}g_{2j}V_{2}+2\sum_k g_{ik}g_{kj} V_\phi \,, \\
 \Delta m_{ij}^2  &=&  \left(g_{i2}g_{2j} + 2\sum_k g_{ik}g_{kj}\right) \frac{T^2}{24} \, , \\
\delta m_{ij} &=& m_{2} \left(\frac{g_{ij}g_{22}}{m_\phi^2(T)}\right)  \frac{T^2}{6} 
\eea
with  $m_\phi^2(T) = m_\phi^2 + \left(\sum_i g_{i2} g_{2i}\right)T^2/3 $.
Here, $V_{ij}$ represents the vector-like thermal potential, $\Delta m_{ij}^2$ denotes the thermal mass-squared correction arising from the momentum-dependent loop diagrams, and $\delta m_{ij}$ is the scalar mass correction induced by the tadpole diagram.

To make contact with the usual two-state oscillation description, we next convert Eq.~\eqref{eq:DispersionPotential}, which is linear in energy and momentum, into an effective mass-squared equation.
Defining the thermally corrected scalar mass matrix
\bea
M_{ij}
\equiv
m_i\delta_{ij}+\delta m_{ij}\, ,
\eea
and squaring Eq.~\eqref{eq:DispersionPotential}, we obtain, to leading order in the thermal corrections and in the relativistic regime, $E^2\approx |\vec p|^2\gg m_{i}^2\,,V_{ij}\,,\Delta m_{ij}^2\,,\delta m_{ij}^2\,,m_i \delta m_{jk}$, the effective Klein-Gordon-like equation
\bea
  \left[ E^2 -|\vec p|^2\right] \delta_{ij} N_j \approx  \left[  \Delta m^2_{ij}+(M^2)_{ij}\right] N_j \,
  \label{eq:EffectiveKG}
\eea
Here, $(M^2)_{ij}$ denotes the matrix product, $(M^2)_{ij}=\sum_k M_{ik}M_{kj}$.
Note that in Eq.~\eqref{eq:EffectiveKG}, the potential $V_{ij}$ drops out of the effective mass-squared matrix.
This is the matrix generalization of the corresponding cancellation in the single-particle dispersion relation~\cite{Quimbay:1995jn}.

As a result, the Hamiltonian for the two-state system of $N_{1\,,2}$ reads
\bea
 {\hat H}\approx {1\over 2|\vec p|}
    \begin{pmatrix} \Delta m^2_{11} + (M^2)_{11} ~&~ \Delta m^2_{12} + (M^2)_{12} \\
    \Delta m^2_{12} + (M^2)_{12} ~&~ \Delta m^2_{22} + (M^2)_{22}
    \end{pmatrix} \,.
\label{eq:Mixing}
\eea
For real and symmetric couplings, the off-diagonal elements satisfy $\hat H_{12}=\hat H_{21}$.
%
The diagonal entries of Eq.~\eqref{eq:Mixing} determine the in-medium level splitting between $N_1$ and $N_2$, 
\bea
{\hat H}_{11}-{\hat H}_{22}={1\over 2 |\vec p|} \left[ \Delta m^2_{11}+(m_1+\delta m_{11})^2
-\Delta m^2_{22} -(m_2+\delta m_{22})^2\right] ,
\eea
while the off-diagonal element
\bea
{\hat H}_{12}
=
\frac{1}{2|\vec p|}
\left[
\Delta m_{12}^2
+
(m_1+m_2+\delta m_{11}+\delta m_{22})\delta m_{12}
\right]
\eea
governs their in-medium conversion.
Both contributions are generated by interactions with the thermal background and therefore need not be associated with any vacuum mass mixing.
In this sense, the thermal plasma provides an effective bridge between $N_1$ and $N_2$: a thermal population of $N_2$ can be converted into the otherwise feebly interacting $N_1$ state through the induced off-diagonal Hamiltonian.
As the temperature evolves, the diagonal level splitting can pass through zero, leading to a resonant enhancement of this conversion.




\section{Thermal Production of Sterile Neutrino Dark Matters}
\label{sec:darkmatter}

\subsection{Resonance transition}

We now examine the production of $N_1$ sterile-neutrino dark matter through its effective mixing with the thermal $N_2$ state, as described in the previous section.
The in-medium mixing becomes maximal when the two diagonal entries of the effective oscillation Hamiltonian become degenerate, ${\hat H}_{11}={\hat H}_{22}$, at a high temperature $T\gg m_{1,2}$.

Recall that $g_{22}$ is taken to be sufficiently large to maintain $N_2$ in thermal equilibrium.
By contrast, the off-diagonal coupling $g_{12}$ is assumed to be much smaller than $g_{22}$, so that $N_1$ remains out of equilibrium and is produced non-thermally.
The self-coupling $g_{11}$ is taken to be negligible throughout our discussion.
Under the hierarchy $g_{22} \gg g_{12} \gg g_{11}$,the resonance condition can be satisfied when $N_1$ is heavier than $N_2$, namely $m_1 > m_2$.

Under the same coupling hierarchy, the thermally corrected scalar mass is approximately given by  $m_\phi^2 (T) = m_\phi^2 + {1\over3} g_{22}^2 T^2$.
Using this thermal mass and expanding the resonance condition to first order in $m_2^2/m_1^2$, we find the resonance temperature
\bea
T_{\rm res} \approx \frac{\sqrt{8}\,m_{1}}{g_{22}}\left( 1-{1\over2} \left( 1+\frac{\frac{4}{3}\frac{m_1^2}{m_\phi^2}}{1+\frac{8}{3}\frac{m_1^2}{ m_\phi^2}}\right)^2\frac{m_{2}^2}{m_{1}^2} \right) \, .
\label{eq:Tres}
\eea
As we will discuss below, the regime $m_\phi\gtrsim m_1$ naturally overlaps with the parameter space in which dark matter is stable and resonant conversion dominates over competing production processes.
As a concrete benchmark, consider $m_1=1\,{\rm TeV}$, $m_2=350\,{\rm GeV}$, $m_\phi=1.2\,{\rm TeV}$, and $g_{22}=10^{-2}$, which satisfies the relevant conditions discussed below.
For this benchmark, Eq.~\eqref{eq:Tres} gives $T_{\rm res}\simeq 2.5\times10^5\,{\rm GeV}$, in good agreement with the numerical solution of the full resonance condition with a difference of less than $1\%$.

\subsection{Production rate}
\label{sec:boltzmann}

In our two-state sterile-neutrino system, $N_2$ is assumed to remain in thermal equilibrium through frequent collisions with the thermal bath.
These collisions continuously damp the coherence between $N_1$ and $N_2$, so that the conversion process cannot, in general, be described as a fully coherent two-state oscillation.
The appropriate framework is therefore the density-matrix formalism~\cite{Stodolsky:1986dx,Sigl:1993ctk,Vlasenko:2013fja}, which consistently describes both the occupation numbers and the coherence between the two states.
For a given momentum mode with relativistic energy $\omega\simeq |\vec p|$, we define the density matrix as
\bea \label{eq:rhoij}
\rho_{ij} \equiv \left\langle a_j^\dagger a_i\right\rangle = {\rm Tr}\left[a_j^\dagger a_i \hat{\rho}\right]\,,
\eea
where $a_i^{(\dagger)}$ is the annihilation (creation) operator for $N_i$, and $\hat{\rho}$ denotes the many-body density operator.
The diagonal components, $\rho_{11}$ and $\rho_{22}$, correspond to the occupation numbers of $N_1$ and $N_2$, respectively, while the off-diagonal components, $\rho_{12}$ and $\rho_{21}=\rho_{12}^*$, account for the coherence between the two states.

The evolution of the one-particle density matrix follows from that of the many-body density operator $\hat{\rho}$.
Including the incoherent interactions with the thermal bath, we write its evolution in Lindblad form as
\bea \label{eq:Lindblad}
\frac{d}{dt}\hat{\rho}
= -i\left[\widehat{\mathcal H}\,,\hat{\rho}\right] + \left(\hat{A} \hat{\rho}\hat{A}^\dagger - \frac{1}{2}\left\{\hat{A}^\dagger\hat{A}\,, \hat{\rho}\right\}\right) + \left(\hat{P} \hat{\rho}\hat{P}^\dagger - \frac{1}{2}\left\{\hat{P}^\dagger\hat{P}\,, \hat{\rho}\right\}\right) \, .
\eea
Here, $\widehat{\mathcal H}$ denotes the Hamiltonian operator governing $N_1$--$N_2$ oscillations, while $\hat{A}$ and $\hat{P}$ are the jump operators describing absorption and production in the thermal medium, respectively.
In the relativistic limit, the part of the Hamiltonian relevant for oscillations can be written as
\bea
\widehat{\mathcal H} = \sum_{ij}a_i^\dagger {\mathcal H}_{ij} a_j
\eea
with the Hamiltonian components $\hat H_{ij}$ given in Eq.~(\ref{eq:Mixing}).

Since $N_2$ is kept in thermal equilibrium through frequent collisions with the thermal bath, the relevant jump operators are
\bea
\hat{A} =\sqrt{A_2}a_2 \,, \quad \hat{P} =\sqrt{P_2}a_2^\dagger \,,
\eea
where $A_2$ and $P_2$ denote the absorption and production rates of  $N_2$  with energy $\omega$, respectively. 
Detailed balance in thermal equilibrium yields $P_2 = A_2e^{-\omega/T}$.

Using the Hamiltonian and jump operators above in Eq.~\eqref{eq:Lindblad}, we obtain the evolution equation for the one-particle density matrix
\bea 
\frac{d}{dt}\rho_{ij} 
= -i \left[{\mathcal H}\,,\rho\right]_{ij} - \frac{1}{2}\left\{\Gamma_{\rm abs}\,,\rho\right\}_{ij} + \frac{1}{2}\left\{\Gamma_{\rm pro}\,,1-\rho\right\}_{ij} \,.
\label{eq:drhodt}
\eea
The factor $(\mathbf{1}-\rho)$ in the production term accounts for Pauli blocking, since $N_i$ are fermions.
In the $(N_1,N_2)$ basis, the absorption and production matrices read
\bea
\Gamma_{\rm abs} = \left(
\begin{tabular}{cc}
    $0$ & $0$ \\
    $0$ & $A_2$
\end{tabular}
\right)\,,\quad
\Gamma_{\rm pro} = \left(
\begin{tabular}{cc}
    $0$ & $0$ \\
    $0$ & $P_2$
\end{tabular}
\right)\,.
\eea
We neglect direct absorption and production of $N_1$, since these processes are suppressed by the feeble couplings involving $N_1$ and are much slower than the corresponding $N_2$ processes in the parameter region of interest.

Throughout the following analysis, we use the dimensionless comoving momentum variable
\bea
x
\equiv
\frac{|\vec p|}{T}\, ,
\eea
which remains constant along free propagation as long as $g_{*s}$ varies negligibly.
Introducing further
\bea
z
\equiv
\frac{m_1}{T}\, ,
\eea
we have $\omega \approx |\vec p|\, = m_1 x/z$
For approximately constant $g_{*s}$, the temperature evolution during radiation domination gives the Hubble parameter relation $H= z^{-1} dz/dt$, where $H$ denotes the Hubble expansion rate.
Eq.~~\eqref{eq:drhodt} can then be rewritten in component form as~\cite{Redondo:2013lna}
\beq
\begin{split}
z {d\rho_{11}\over dz} &= i \delta (\rho_{12}-\rho_{21}) \\
z {d\rho_{12}\over dz} &= -(D+i \Delta)\rho_{12}+i\delta (\rho_{11}-\rho_{22}) \\
z {d\rho_{22}\over dz} &= -i \delta (\rho_{12}-\rho_{21})- 2D (\rho_{22}-f_{\rm eq})
\end{split}
 \label{eq:drhoij}
\eeq
where 
\bea
\delta
\equiv
\frac{\mathcal H_{12}}{H}\, ,
\qquad
\Delta
\equiv
\frac{\mathcal H_{11}-\mathcal H_{22}}{H}\, ,
\qquad
2D
\equiv
\frac{A_2+P_2}{H}
=
\frac{(1+e^x)P_2}{H}\, ,
\eea
and $f_{\rm eq}=1/(1+e^{x})$.
Here, $\delta$ characterizes the in-medium mixing, $\Delta$ the level splitting between the two sterile states, and $D$ the collisional damping rate normalized to the Hubble expansion rate.

In the parameter region of interest, the mixing rate is much smaller than the collisional damping rate, $\delta/D \sim g_{12}/x g_{22}^3 \ll1 $, corresponding to the strong-damping, or ``{\it quantum Zeno}" regime.
At the same time, the damping parameter around the resonance $z_{\rm res}\sim g_{22}/\sqrt{8}$ is estimated as $D\sim (g_{22}^5/20\pi \sqrt{g_*})(m_{\rm pl}/m_1)(z/z_{\rm res}) \gg1$, where $m_{\rm pl}=1.22\times10^{19}\,{\rm GeV}$ denotes the non-reduced Planck mass.
This condition ensures that $N_2$ relaxes toward thermal equilibrium much faster than the Hubble expansion, so that $\rho_{22}\approx f_{\rm eq}$ throughout the resonant transition.

Provided that collisional damping acts sufficiently rapidly, the off-diagonal component follows its instantaneous quasi-stationary solution,
\bea
\rho_{12}
\simeq
\frac{i\delta}{D+i\Delta}
\left(
\rho_{11}-\rho_{22}
\right)\, .
\label{eq:rho12stationary}
\eea
As we will discuss below, the validity of this approximation requires an appropriate hierarchy between the damping timescale and the timescale over which the background quantities vary.
Importantly, however, in the perturbative weak-conversion regime, this condition is not necessarily required for obtaining the leading conversion probability integrated over the resonance as described later.

Substituting Eq.~\eqref{eq:rho12stationary} into the first equation of Eq.~\eqref{eq:drhoij}, and using $\rho_{22}\simeq f_{\rm eq}$, gives
\bea
z\frac{d\rho_{11}}{dz} = -
\frac{2\delta^2D}
{\Delta^2+D^2}
\left(
\rho_{11}-f_{\rm eq}
\right)\, .
\label{eq:rho11evolution}
\eea
Within this quasi-stationary description, an initially negligible occupation number, $\rho_{11}(z_0,x)=0$, grows monotonically toward $f_{\rm eq}(x)$ and does not exceed the equilibrium occupation number of the thermal $N_2$ state.


The evolution equation in Eq.~\eqref{eq:rho11evolution} is solved by
\beq
\begin{split}
\rho_{11}(x) &= \left(1-e^{-I(x)}\right) f_{\rm eq}(x), \\
I(x) &\equiv \int^{z_\infty}_{z_0} {2\delta^2 D\over \Delta^2 + D^2} {dz\over z} \,.
\end{split}
\label{eq:rho11}
\eeq
Here, $z_0$ and $z_\infty$ denote points sufficiently before and
after the resonance, respectively.
Far away from the level crossing, $|\Delta|\gg \delta,D$, such that the in-medium mixing is strongly suppressed and the contribution to the integral becomes negligible.
The precise choices of $z_0$ and $z_\infty$ are therefore irrelevant as long as they lie sufficiently far from the resonant region.
Since $x=|\vec p|/T$ is fixed, $f_{\rm eq}(x)$ remains constant during the evolution.

Around the resonance, where $\Delta(z_{\rm res})\simeq 0$, the level splitting can be expanded linearly as $\Delta\simeq (d\Delta/d\ln z_{\rm res})(\ln z - \ln z_{\rm res})$.
If the resonance is sufficiently narrow that $\delta$, $D$, and the remaining background quantities vary negligibly across the resonant interval, the integral is dominated by the vicinity of
$z=z_{\rm res}$.
The integration range can then be effectively extended well beyond the resonance on both sides, yielding
\bea 
I_{\rm res} = 2\pi \left.{\delta^2 \over |d\Delta/d\ln z|}\right|_{\rm res} \, .
\label{eq:Ires}
\eea
Here, the characteristic width of the resonance in $\ln z$ is $D/(d\Delta/d\ln z_{\rm res})$, which is much smaller than unity in the narrow-resonance limit.

An interesting feature of Eq.~\eqref{eq:Ires} is that the integrated resonant conversion becomes independent of the damping rate $D$.
This can be understood directly from the resonance profile in Eq.~\eqref{eq:rho11evolution}.
At $\Delta=0$, its peak height scales as $D^{-1}$, whereas its width in the level splitting scales as $D$.
Consequently, increasing the collision rate lowers the instantaneous conversion rate at the center of the resonance but broadens the interval over which the conversion takes place.
These two effects compensate each other upon integration,
leaving the total conversion insensitive to the detailed damping rate at leading order.
This cancellation holds provided that the resonance can be treated as approximately linear and the other quantities vary slowly across the resonant interval.

Interestingly, within the quasi-stationary description, the conversion fraction $\rho_{11}/f_{\rm eq}$ has the same functional form as the transition probability in the coherent Landau--Zener problem.
Introducing the Landau--Zener adiabaticity parameter
\bea
\gamma_{\rm LZ}
\equiv
\left.
\frac{
4{\mathcal H}_{12}^2
}{
\left|
d({\mathcal H}_{11}-{\mathcal H}_{22})/dt
\right|
}
\right|_{\rm res}\, ,
\eea
we find $I_{\rm res} = \frac{\pi}{2}\gamma_{\rm LZ}$.
The conversion fraction therefore becomes
\bea
\frac{\rho_{11}}{f_{\rm eq}}
=
1-
\exp\left(
-\frac{\pi}{2}\gamma_{\rm LZ}
\right)\, ,
\eea
which coincides with the coherent Landau--Zener transition probability.

This correspondence should not be interpreted as implying that the microscopic evolution in our system is coherent.
Rather, within the quasi-stationary approximation, collisional damping modifies the local profile of the resonant conversion rate while leaving its integrated strength unchanged as described above.
For arbitrary $I_{\rm res}$, the correspondence relies on the validity of the quasi-stationary treatment, the maintenance of $\rho_{22}\simeq f_{\rm eq}$, and an approximately linear level crossing with slowly varying mixing and damping parameters.
In the perturbative weak-conversion regime, $I_{\rm res}\ll1$, however, the leading integrated result remains valid even if the local quasi-stationary approximation is relaxed, provided that the level crossing remains approximately linear and the relevant background quantities vary slowly over the transition interval.

Since $I_{\rm res}$ is inversely proportional to the dimensionless momentum $x=|\vec p|/T$, we parameterize its momentum dependence as
\bea
I_{\rm res}(x)
\equiv
2\pi\frac{x_c}{x}\, ,
\eea
where
\bea
x_c
\simeq
 6.7 \times 10^{-3}\,
\frac{c_r}{\sqrt{g_*}}
g_{12}^2g_{22}
\frac{m_{\rm pl}}{m_1}\, .
\label{eq:xc}
\eea
Here, $c_r$ is a dimensionless kinematic function of the mass ratios $m_2/m_1$ and $m_\phi/m_1$, and is typically of order unity in the parameter region of interest.
For the benchmark given above, we find $c_r\simeq1.52$.
More generally, $c_r$ remains of order unity for moderately
hierarchical $m_1$ and $m_2$, while it can become larger as the
quasi-degenerate limit $m_2\simeq m_1$ is approached.

For the freeze-in production considered here, the level crossing is non-adiabatic and the conversion probability is small, $I_{\rm res}(x)\ll1$.
The final $N_1$ occupation number can therefore be approximated as
\bea
\rho_{11}(x)
\simeq
I_{\rm res}(x)f_{\rm eq}(x)
=
2\pi\frac{x_c}{x}f_{\rm eq}(x)\, .
\eea
The additional $1/x$ dependence implies that the resulting nonthermal spectrum is biased toward lower momenta compared with a thermal relativistic Fermi--Dirac distribution.
In particular, its average dimensionless momentum is $\langle p/T\rangle_{N_1} = 18\zeta(3)/\pi^2\simeq 2.19$, to be compared with $\langle p/T\rangle_{\rm FD}=7\pi^4/180\zeta(3)\simeq3.15$ for a thermal relativistic fermion.
Thus, resonant freeze-in produces a characteristically colder momentum distribution.

Integrating over momentum, the resulting $N_1$ number density is
\bea
n_1 =
g_{N_1}
\int
\frac{d^3\vec p}{(2\pi)^3}
\rho_{11}(x) \simeq
g_{N_1}
\frac{x_c}{\pi}
T^3
\int_0^\infty
dx\,
x f_{\rm eq}(x) =
g_{N_1}
\frac{\pi}{12}
x_cT^3\, ,
\eea
where $g_{N_1}=2$ denotes the spin multiplicity of $N_1$.
The corresponding comoving abundance reads
\bea
Y_1
\equiv
\frac{n_1}{s}
=
\frac{15}{4\pi g_{*s}}x_c
\simeq
\frac{
0.8\times10^{-2}\,c_r
}{
g_{*s}\sqrt{g_*}
}
g_{12}^2g_{22}
\frac{m_{\rm pl}}{m_1}\, ,
\label{eq:Yres}
\eea
where $s=(2\pi^2/45)g_{*s}T^3$ is the entropy density.


\subsection{Conditions for viable resonant production}
\label{sec:conditions}

The observed comoving dark matter energy density is $\rho_{\rm DM}^{\rm obs}/s\simeq
0.44\,{\rm eV}$.
Using the resonant abundance obtained in Eq.~\eqref{eq:Yres}, the condition $m_1Y_1=\rho_{\rm DM}^{\rm obs}/s$ results in
\bea
g_{12}
\simeq
2.2\times10^{-11}
\left(
\frac{10^{-2}}
{c_r g_{22}}
\right)^{1/2}
\left(
\frac{g_{*s}}{106.75}
\right)^{1/2}
\left(
\frac{g_*}{106.75}
\right)^{1/4}\, .
\label{eq:reliccondition}
\eea
A notable feature of resonant production is that the comoving dark matter energy density $m_1Y_1$ is independent of $m_1$, apart from the mass-ratio dependence contained in $c_r$ and possible variations of $g_*$ and $g_{*s}$.

\subsubsection*{Dynamical consistency of resonant production}

Several dynamical conditions must be satisfied for the resonant-production scenario considered above.
First, $N_2$ must remain in thermal equilibrium throughout the resonance.
This requires its relaxation rate to be much larger than the Hubble expansion rate
\bea
D_{\rm res}
\gg
1\, .
\label{eq:N2equilibrium}
\eea
Second, the mixing rate must remain smaller than the collisional damping rate
\bea
\frac{\delta_{\rm res}}
{D_{\rm res}}
\ll
1\, ,
\label{eq:quantumZeno}
\eea
which places the system in the strong-damping, or quantum-Zeno regime.
Using the estimates given below Eq.~\eqref{eq:drhoij}, the equilibrium condition therefore gives the approximate upper bound
\bea
m_1
\ll
\frac{g_{22}^5}
{20\pi\sqrt{g_*}}
m_{\rm pl}
\simeq
1.9\times10^6\,{\rm GeV}
\left(
\frac{g_{22}}{10^{-2}}
\right)^5
\left(
\frac{106.75}{g_*}
\right)^{1/2}\, .
\label{eq:m1upper}
\eea
If we further restrict our analysis to $m_1\gtrsim v_{\rm EW}$, as adopted in the following, the existence of a viable mass interval requires approximately
\bea
g_{22}
\gtrsim
{\cal O}(10^{-3})\, .
\eea
The condition $m_1\gtrsim v_{\rm EW}$ is a sufficient choice ensuring that the high-temperature treatment is applicable; more generally, it is the resonance temperature $T_{\rm res}$ that should lie in the regime where the assumed Higgs-portal thermalization is valid.


The freeze-in calculation also assumes perturbatively small conversion for the momentum modes that dominate the abundance,
\bea
I_{\rm res}(x)
=
2\pi\frac{x_c}{x}
\ll
1
\qquad
{\rm for}
\qquad
x={\cal O}(1)\, .
\label{eq:weakconversion}
\eea
Equivalently, it is sufficient that $x_c\ll1$.
Using the observed relic abundance, one finds
\bea
x_c
=
\frac{4\pi g_{*s}}{15}
\frac{\rho_{\rm DM}^{\rm obs}/s}{m_1}
\simeq
1.6\times10^{-10}
\left(
\frac{g_{*s}}{106.75}
\right)
\left(
\frac{m_1}{v_{\rm EW}}
\right)^{-1}\, ,
\eea
so that weak conversion is comfortably satisfied throughout the mass range considered here.
Modes with extremely small $x$ may formally have $I_{\rm res}(x)\gtrsim1$, but their phase-space volume is negligible when $x_c\ll1$.

The cosmological thermal history must also reach the resonance condition $T_{\rm RH} \gtrsim T_{\rm res}$, where $T_{\rm RH}$ denotes the reheating temperature.
In addition, both $N_2$ and $\phi$ must be thermally populated at $T_{\rm res}$, as assumed in the previous section.

For completeness, the local quasi-stationary replacement in Eq.~\eqref{eq:rho12stationary} requires the relaxation timescale of the off-diagonal component to be shorter than the timescale over which its instantaneous stationary value changes.
Near the resonance, this condition can be written as
\bea
D_{\rm res}^2
\gg
\left|
\frac{d\Delta}{d\ln z}
\right|_{\rm res}\, .
\label{eq:quasistationarycondition}
\eea
We emphasize, however, that this condition is required only for the local quasi-stationary description and for the corresponding all-orders solution in Eq.~\eqref{eq:rho11}.
In the perturbative weak-conversion regime relevant for freeze-in, i.e., $I_{\rm res} \ll 1$, the leading conversion probability integrated over the resonance remains valid even when the local quasi-stationary approximation is relaxed as discussed above.
We therefore regard Eq.~\eqref{eq:quasistationarycondition} as a consistency condition for the local analytic treatment rather than as an independent phenomenological bound on the leading freeze-in abundance.

For the analytic treatment of the resonant integral, we instead require the level crossing to be sufficiently narrow that the mixing, damping rate, and equilibrium source vary negligibly over the transition interval.
A sufficient condition is
\bea
\frac{D_{\rm res}}
{
\left|
d\Delta/d\ln z
\right|_{\rm res}
}
\ll
1\, .
\label{eq:narrowresonancecondition}
\eea
Using $z_{\rm res}\simeq g_{22}/\sqrt{8}$ and the relativistic expressions for the level splitting and damping rate, this ratio is estimated as $D_{\rm res}/\left|d\Delta/d\ln z\right|_{\rm res}\sim x\,g_{22}^2$.
Thus, for the thermally relevant momenta $x={\cal O}(1)$, the narrow-resonance approximation is well justified for $g_{22}\ll1$.



\subsubsection*{Dark matter stability}

Since $N_1$ is the dark matter candidate, its lifetime must exceed the age of the Universe,
\bea
t_U
\simeq
4.4\times10^{17}\,{\rm sec}\, .
\eea
If the two-body decay $N_1 \rightarrow N_2+\phi$ is kinematically allowed, namely if $m_1>m_2+m_\phi$, its decay rate away from threshold is approximately $\Gamma_{N_1\rightarrow N_2\phi} \simeq (g_{12}^2/16\pi) m_1$
up to the corresponding two-body phase-space factor.
For the value of $g_{12}$ required by Eq.~\eqref{eq:reliccondition}, this rate is much larger than $t_U^{-1}$.
The decay must therefore be forbidden kinematically, requiring
\bea
m_1
<
m_2+m_\phi\, .
\label{eq:twobodystability}
\eea

A further decay channel is mediated by an off-shell $\phi$, $N_1\rightarrow N_2+\phi^* \rightarrow N_2+N_2+N_2$, which is kinematically allowed for $m_1>3m_2$.
In the heavy-mediator limit, where the virtuality carried by $\phi^*$ is much smaller than $m_\phi^2$, the decay rate can be estimated as $\Gamma_{N_1\rightarrow3N_2} \sim (g_{12}^2g_{22}^2/1536\pi^3)m_1^5/m_\phi^4$ up to order-one factors associated with the identical Majorana final states and their interference.
Requiring $\Gamma_{N_1\rightarrow3N_2}<t_U^{-1}$ gives
\bea
m_\phi
\gtrsim
\left(
\frac{
g_{12}^2g_{22}^2m_1^5t_U
}{
1536\pi^3
}
\right)^{1/4}\, .
\label{eq:threebodystability}
\eea
At the same time, $\phi$ should remain thermally populated around the resonance.
In addition to requiring that its Higgs-portal interaction maintain equilibrium, a simple sufficient condition for avoiding strong Boltzmann suppression is
\bea
m_\phi
\lesssim
T_{\rm res}\, .
\label{eq:phithermalpopulation}
\eea
Combining this condition with Eq.~\eqref{eq:threebodystability} can strongly restrict the viable parameter region when the three-body decay is open.
We therefore focus on the simpler kinematic regime
\bea
m_1
<
3m_2\, ,
\label{eq:threebodykinematic}
\eea
for which the three-body decay is forbidden.

\subsubsection*{Direct production through scattering}

For resonant conversion to provide the dominant contribution to the $N_1$ abundance, direct freeze-in processes induced by $g_{12}$ must remain subdominant~\cite{Hall:2009bx}.
Representative $2\rightarrow2$ processes include $N_2 +\phi\rightarrow N_1 +\phi$, $\phi+\phi \rightarrow N_2+N_1$, and $N_2 + N_2 \rightarrow N_1 +N_2$.
Each of these processes contains one insertion of the feeble off-diagonal coupling $g_{12}$ and one insertion of the larger coupling $g_{22}$.
In the relativistic regime, the corresponding thermally averaged cross section scales as $\left<\sigma_{12}v\right>\sim g_{12}^2g_{22}^2/16\pi T^2$.
The associated freeze-in abundance is parametrically
\bea
\left.Y_1\right|_{\rm sca}
\simeq
4\times 10^{-5}c_{\rm s}
\frac{
g_{12}^2g_{22}^2
}{
g_{*s}\sqrt{g_*}
}
\frac{
m_{\rm pl}
}{
T_{\rm IR}
}\, ,
\label{eq:Yscattering}
\eea
where $T_{\rm IR}\sim\max(m_1,m_2,m_\phi)$ denotes the temperature at which the reaction becomes Boltzmann suppressed, and $c_{\rm s}$ is a $\mathcal{O}(1)$ constant containing the thermal phase-space, angular, and multiplicity factors.

Comparing Eq.~\eqref{eq:Yscattering} with the resonant abundance in Eq.~\eqref{eq:Yres}, we find
\bea
\frac{
\left.Y_1\right|_{\rm sca}
}{
\left.Y_1\right|_{\rm mix}
}
\sim0.5\times10^{-2}
\frac{c_{\rm s}}{c_r} g_{22} \frac{m_1}{T_{\rm IR}}\, .
\label{eq:scatteringratio}
\eea
Thus, the direct scattering contribution is suppressed relative to resonant conversion by an additional power of $g_{22}$, together with the numerical suppression arising from the $2\rightarrow2$ phase-space integral.
For perturbative $g_{22}$ and comparable masses, these processes are therefore expected to be subdominant.
We use Eq.~\eqref{eq:scatteringratio} only to establish the parametric hierarchy; a precise numerical comparison requires evaluating the full matrix elements and thermal phase-space integrals.

Processes involving two insertions of $g_{12}$, such as
$N_2N_2\rightarrow N_1N_1$, are further suppressed by additional powers of the feeble off-diagonal coupling.
The decay $\phi\rightarrow N_1N_1$ is negligible under our assumption $g_{11}\ll g_{12}$, while $N_2\rightarrow N_1+\phi$ is forbidden by the mass hierarchy $m_1>m_2$.
When evaluating the scattering contribution, any real-intermediate-state contribution associated with an on-shell $\phi$ should be subtracted and included instead in the decay contribution discussed below.

\subsubsection*{Direct production from scalar decay}

The potentially most important competing production channel is $\phi \rightarrow N_1+N_2$\, which is kinematically allowed when
\bea
m_\phi > m_1+m_2\, .
\label{eq:phidecayopen}
\eea
We parameterize the rest-frame decay width as
\bea
\Gamma_\phi
=
\frac{g_{12}^2m_\phi}{8\pi}
{\cal F}_\phi\, ,
\label{eq:phidecaywidth}
\eea
where the phase-space function for a scalar Yukawa interaction is
\bea
{\cal F}_\phi
=
\lambda^{1/2}
\left(
1,
\frac{m_1^2}{m_\phi^2},
\frac{m_2^2}{m_\phi^2}
\right)
\left[
1-
\left(
\frac{m_1+m_2}{m_\phi}
\right)^2
\right]\, .
\label{eq:phasespacefactor}
\eea
Here,
\bea
\lambda(a,b,c)
=
a^2+b^2+c^2-2ab-2ac-2bc
\eea
is the K\"all\'en function.

For a thermal population of $\phi$, the production rate per unit volume is
\bea
\left.\gamma_1\right|_{\rm dec}
=
n_\phi^{\rm eq}
\Gamma_\phi
\frac{
K_1(m_\phi/T)
}{
K_2(m_\phi/T)
}\, .
\label{eq:decayreactiondensity}
\eea
Performing the thermal integral in the Maxwell--Boltzmann approximation gives the standard freeze-in yield
\bea
\left.Y_1\right|_{\rm dec}
\simeq
\frac{135g_\phi}{8\pi^3(1.66) g_{*s}\sqrt{g_*}}\frac{m_{\rm pl}\Gamma_\phi}{m_\phi^2}
=
\frac{135}{64\pi^4(1.66)}
\frac{g_\phi{\cal F}_\phi}{g_{*s}\sqrt{g_*}}g_{12}^2
\frac{m_{\rm pl}}{m_\phi}\, .
\label{eq:Ydecay}
\eea
Here, $g_\phi$ denotes the number of thermally populated scalar degrees of freedom; for a real scalar, $g_\phi=1$.

Comparing this result with the resonant abundance in Eq.~\eqref{eq:Yres}, we obtain
\bea
\frac{
\left.Y_1\right|_{\rm mix}
}{
\left.Y_1\right|_{\rm dec}
}
\simeq
0.6\,
\frac{
c_r
}{
g_\phi{\cal F}_\phi
}
g_{22}
\frac{
m_\phi
}{
m_1
}
\simeq
1.7\,
\frac{
c_r
}{
g_\phi{\cal F}_\phi
}
\frac{
m_\phi
}{
T_{\rm res}
}\, .
\label{eq:decayratio}
\eea
In the second relation, we use the leading-order resonance temperature $T_{\rm res}
\simeq
\sqrt{8}\,m_1/g_{22}$.
Therefore, when $\phi\rightarrow N_1+N_2$ is kinematically open, resonant conversion dominates over direct decay production provided that
\bea
m_\phi\gtrsim 0.6\,
\frac{g_\phi{\cal F}_\phi}{c_r}T_{\rm res}\, .
\label{eq:decaydominance}
\eea
This numerical criterion is subject to order-one corrections associated with the Maxwell--Boltzmann approximation, quantum-statistical effects, and the counting of thermally populated scalar degrees of freedom.
Moreover, near the decay threshold ${\cal F}_\phi\ll1$, so that direct decay production is strongly phase-space suppressed and the above condition becomes correspondingly weaker.

Together with the requirement that $\phi$ remain thermally populated at the resonance, $m_\phi\lesssim T_{\rm res}$, there can exist a finite region in which the scalar decay is kinematically open but remains subdominant,
\bea
0.6\,
\frac{
g_\phi{\cal F}_\phi
}{
c_r
}
T_{\rm res}
\lesssim
m_\phi
\lesssim
T_{\rm res}\, .
\eea
Since ${\cal F}_\phi$ itself depends on $m_\phi$, this inequality should be understood as an implicit condition.
The window exists only when its lower boundary lies below $T_{\rm res}$.

If instead
\bea
m_\phi
<
m_1+m_2\, ,
\eea
the decay $\phi\rightarrow N_1+N_2$ is kinematically forbidden, and no additional abundance constraint arises from this channel.
Combining this condition with the two-body stability requirement in Eq.~\eqref{eq:twobodystability}, and recalling that $m_1>m_2$, gives
\bea
m_1-m_2
<
m_\phi
<
m_1+m_2\, .
\label{eq:masswindow}
\eea
Together with the three-body stability condition $m_1<3m_2$ in Eq.~\eqref{eq:threebodykinematic}, this defines a simple kinematic region in which $N_1$ is cosmologically stable and its abundance can be generated predominantly through resonant in-medium conversion.

\section{Discussions and Conclusion}
\label{sec:conc}

We have proposed and analyzed a resonant production mechanism for sterile-neutrino dark matter based entirely on thermal mixing within a sterile sector. The setup contains two sterile neutrinos, $N_1$ and $N_2$, coupled to a singlet scalar $\phi$. The state $N_1$ is the dark matter candidate, while $N_2$ is kept in thermal equilibrium through its interactions with the sterile sector and the primordial plasma. In contrast to conventional sterile-neutrino production, neither the mixing responsible for dark matter production nor the resonance itself relies on appreciable active--sterile mixing.

The key ingredient is the modification of sterile-neutrino propagation in the thermal medium. Thermal self-energies generated by the $N_2$ and $\phi$ backgrounds induce temperature-dependent diagonal and off-diagonal contributions to the sterile-neutrino effective mass-squared matrix. The vector-like thermal potential cancels from the relativistic dispersion relation, whereas the remaining thermal mass corrections generate an evolving level splitting and an effective $N_1$--$N_2$ mixing. For $m_1>m_2$, the two in-medium states can
therefore undergo a level crossing as the Universe cools, providing a resonant conversion channel from the thermal $N_2$ population into $N_1$.

Because $N_2$ continuously interacts with the thermal bath, the conversion process cannot, in general, be described as an isolated coherent two-state oscillation.
We have therefore formulated the dynamics using a density-matrix approach that incorporates coherent $N_1$--$N_2$ conversion together with production, absorption, and collisional decoherence of $N_2$.
In the parameter region relevant for freeze-in, $N_2$ remains close to thermal equilibrium and the system lies in the strong-damping, or quantum-Zeno, regime.
When the off-diagonal coherence follows its quasi-stationary solution, the resulting production rate exhibits a Lorentzian resonance profile whose peak height decreases with the damping rate while its width increases correspondingly.
For a narrow and approximately linear level crossing, these two effects compensate after integration, rendering the total resonant conversion independent of the damping rate at leading order.
The resulting conversion fraction has the same functional form as the coherent Landau--Zener transition probability.
Importantly, in the perturbative weak-conversion regime relevant for freeze-in, the leading integrated Landau--Zener-like result remains valid even when the local quasi-stationary approximation is relaxed, provided that the level crossing is approximately linear and the relevant background quantities vary slowly across the transition interval.

In the weak-conversion regime, we find
$
Y_1^{\rm mix}\propto
g_{12}^2 g_{22}\frac{M_{\rm Pl}}{m_1},
$
so that the dark matter energy density $m_1Y_1^{\rm mix}$ is approximately independent of the dark matter mass. Requiring the observed relic abundance fixes the off-diagonal coupling at the level
$
g_{12}\sim 10^{-12}/\sqrt{g_{22}},
$
up to the order-one dependence associated with the sterile-sector mass hierarchy. 
We have also examined the conditions required for $N_1$ to remain cosmologically stable. The potentially dangerous two- and three-body decay channels can be kinematically closed, while direct $2\to2$ freeze-in production is parametrically subdominant to resonant conversion. Even when the decay $\phi\to N_1N_2$ is open, there exists a finite region in which its contribution remains smaller than the resonant one.

Our results demonstrate that a thermal sterile sector can provide its own resonant environment for dark matter production.
The mechanism separates the origin of the dark matter abundance from active--sterile mixing, which is responsible for the characteristic observational constraints on conventional sterile-neutrino dark matter scenarios.
It therefore opens a qualitatively different avenue for sterile-neutrino dark matter, particularly for masses above the electroweak scale, where the cosmological abundance is governed primarily by the thermal history and interactions within the sterile sector rather than by direct mixing with active neutrinos.

\paragraph{Acknowledgments.} 
The authors thank Kyu Jung Bae and Yurang Ko for useful discussions.
SY is supported by Basic Science Research Program through the National Research Foundation of Korea(NRF) funded by the Ministry of Education(No. RS-2026-25571203) and by IBS under the project code, IBS-R018-D1.
This work was in part supported by Center of Quantum Cosmo Theoretical Physics (NSFC grant No. 12347103).
\appendix

\bibliographystyle{JHEP}
\bibliography{SNuDM}

\end{document}